\documentclass[aps,pra,twocolumn,showpacs,10pt]{revtex4-1}
\usepackage{amsmath,amssymb,graphicx,stmaryrd}
\usepackage[utf8]{inputenc}
\usepackage[T1]{fontenc}
\usepackage{abraces}

\IfFileExists{newtxtext.sty}
   {\usepackage{newtxtext,newtxmath}}
   {\IfFileExists{stix.sty}
      {\usepackage{stix}}
      {\IfFileExists{mathptmx.sty}
      {\usepackage{mathptmx}}{} } }

\usepackage{textcomp}

\usepackage{bm}
\usepackage{bbold}
\usepackage{multirow}
\usepackage{siunitx}

\newcommand{\bra}[1]{\langle#1\rvert}
\newcommand{\ket}[1]{\lvert#1\rangle}
\newcommand{\overlap}[2]{\langle#1\vert#2\rangle}
\newcommand{\expectation}[1]{\langle#1\rangle}

\newcommand{\linkdoi}[2]{\href{http://dx.doi.org/#2}{#1}}
\newcommand{\linkarxiv}[2]{\href{http://arxiv.org/abs/#2}{#1}}

\usepackage{xcolor}
\definecolor{blue}{rgb}{0.0, 0.38, 0.68}   
\definecolor{red}{rgb}{0.87, 0.09, 0.12}   
\definecolor{green}{rgb}{0.0, 0.50, 0.0}   
\definecolor{brown}{rgb}{0.62, 0.50, 0.44} 

\newcommand{\sztot}{S^z_\text{tot}}

\newcommand{\green}[1]{\textcolor{green}{#1}}
\newcommand{\blue}[1]{\textcolor{blue}{#1}}
\newcommand{\red}[1]{\textcolor{red}{#1}}

\newcommand{\up}{{\bullet}}
\newcommand{\dn}{{\circ}}

\newcommand{\po}{{\makebox[0.5em]{(}}} 
\newcommand{\pc}{{\makebox[0.5em]{)}}} 

\newcommand{\mo}{{\makebox[0.5em]{\red{$\langle$}}}} 
\newcommand{\mc}{{\makebox[0.5em]{\red{$\rangle$}}}} 

\newcommand{\co}{{\makebox[0.5em]{\green{$\lceil$}}}} 
\newcommand{\cc}{{\makebox[0.5em]{\green{$\rceil$}}}} 

\newcommand{\aco}{{\makebox[0.5em]{\blue{$\lfloor$}}}} 
\newcommand{\acc}{{\makebox[0.5em]{\blue{$\rfloor$}}}} 

\pdfoutput=1
\usepackage{color}
\definecolor{LinkColor}{rgb}{0.256,0.439,0.588}
\usepackage{hyperref}
\hypersetup{
   pdfauthor={Khagendra Adhikari and K. S. D. Beach},
   pdftitle={Slow dynamics of the Fredkin spin chain},
   pdfsubject={Quantum Monte Carlo study of the Fredkin spin chain},
   pdfkeywords={Fredkin model} {spin chain} {slow dynamics} {dynamical exponent},
   colorlinks=true,
   citecolor=LinkColor,
   linkcolor=LinkColor,
   urlcolor=LinkColor
   }

\begin{document}

\title{Slow dynamics of the Fredkin spin chain}

\author{Khagendra Adhikari}
\email[Electronic address:\ ]{kadhikar@go.olemiss.edu}
\affiliation{Department of Physics and Astronomy, The University of Mississippi, University, Mississippi 38677, USA}

\author{K. S. D. Beach}
\email[Electronic address:\ ]{kbeach@olemiss.edu}
\affiliation{Department of Physics and Astronomy, The University of Mississippi, University, Mississippi 38677, USA}

\begin{abstract} The dynamical behavior of a quantum many-particle system is
characterized by the lifetime of its excitations. When the system is
perturbed, observables of any non-conserved quantity decay exponentially, but
those of a conserved quantity relax to equilibrium with a power law. Such
processes are associated with a dynamical exponent $z$ that relates the spread
of correlations in space and time. We present numerical results for the
Fredkin model, a quantum spin chain with a three-body interaction term, which
exhibits an unusually large dynamical exponent. We discuss our efforts to
produce a reliable estimate $z \doteq 3.16(1)$ through direct simulation of
the quantum evolution and to explain the slow dynamics in terms of an excited
bond that executes a constrained random walk in Monte Carlo time.
\end{abstract}

\maketitle

\section{Introduction}
\label{SECT:Introduction}

The Fredkin spin chain~\cite{Salberger-arXiv-16} is a fascinating example of a
quantum system that exhibits slow evolution by virtue of dynamical
jamming~\cite{Nussinov-PRB-13,Klich-NatCommun-14,Yang-PNAS-15}. Its
Hamiltonian can be viewed as a sum of operators acting on three adjacent sites
that reconfigure the local spin state in a way that is compatible with a
global conserved quantity, viz., the number of defects in the Dyck word sense.
In practice, this means that the system can be prepared in a state in which
only a microscopic portion of the chain is free to evolve, and the remaining
spins belong to a jammed region that takes a very long time to unwind.

In a typical, translationally invariant quantum magnet, excitations are
characterized by a magnon dispersion relation $\epsilon_k \sim k^z$ with
dynamical exponent $z=1$ or $z=2$ corresponding to antiferromagnet and
ferromagnetic spin waves, respectively. In contrast, the Fredkin model shows
$z \approx 3$. Most intriguing, this large value of the dynamical exponent
occurs in the absence of frustration or explicit competing interactions in the
Hamiltonian.

Since the Fredkin spin chain cannot be mapped to a conformal field theory,
there is no straightforward analytical estimate for $z$, and what we have are
numerical estimates determined by level spectroscopy~\cite{Bravyi-PRL-12,
DellAnna-PRB-16, Chen-PRB-17, Chen-JPA-17, Adhikari-PRB-19}, i.e., from the
scaling of the lowest-lying energy spacing as a function of system size. In a
system of interacting particles, the correlation length $\xi$ defines a length
scale at which differences between microscopic and macroscopic properties are
distinguishable. The dynamical exponent characterizes the powerlaw
relationship between the correlation length and the gap in the spectrum
between the ground state and the lowest-lying excitations. With the onset of
long-range correlations, $\xi$ diverges, up to the point where it is cut off
by the finite system size. Since the Fredkin model is defined on a finite
linear chain segment consisting of $N$ spins, the energy spacing acquires a
power-law scaling, $\Delta \sim \xi^{-z} \sim
N^{-z}$~\cite{Sachdev-Cambridge-11}.

There are three disadvantages to the spectroscopic approach. First, the
particular value of $z$ is sensitive to details of the finite-size scaling
ansatz, especially to the assumptions one must make about subleading
corrections. Second, the approach is inherently black-box and produces little
insight into the physical mechanism that produces the enhanced $z$. Third, it
is subject to the limitations of the numerical technique used to obtain the
energy spectrum. The Fredkin spin chain, for instance, sits in
model parameter space at a special tuning point with a locally maximum
entanglement entropy~\cite{Zhang-JPA-17, Zhang-PNAS-17, Salberger-JSM-17,
Udagawa-JPA-17, Chen-PRB-17,Adhikari-PRB-19}, where the density matrix
renormalization group (DMRG) algorithm has the greatest difficulty converging.

\begin{figure}
\begin{center}
\includegraphics[width = 0.9\columnwidth]{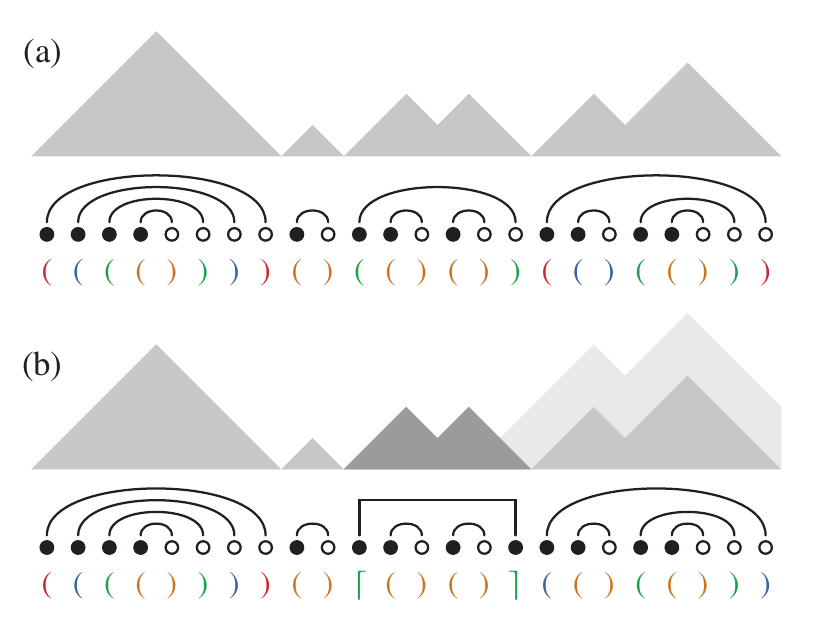}
\end{center}
\caption{\label{FIG:dyck-word} 
(a) A spin configuration that contributes to the ground state is shown as a
nonnegative height profile, a noncrossing bond pattern, and a set of properly
matched parentheses. All three representations are equivalent. (b) A valid
configuration in the $\sztot = 1$ excited state is produced by flipping a spin
at the right edge of a mountain range. We understand this as producing a
single excited (sharp-cornered) bond in the background of conventional
(rounded) bonds.
}
\end{figure}

We adopt an alternative approach based on a faithful, asymptotically exact
simulation of the quantum evolution in Monte Carlo (MC) time and on direct
observation of the behavior of the excited quasiparticles. We devise a
numerical experiment in which an excitation is injected and allowed to
propagate across the system. The time it takes for an excitation to traverse
the chain (moving from left to right for net spin projection $\sztot = 1$
or from right to left for $\sztot = -1$) provides a quantitative measure
of the dynamical exponent, according to~\cite{Hohenberg-RMP-77}
\begin{equation}\label{EQ:DynamicalLaw}
\tau \sim \xi^{z} \sim N^z.
\end{equation}

\begin{table*}[t!]
\caption{\label{TABLE:low-energy-states-N6-N8} 
Normalized wave functions are given for the $N=6$ and $N=8$ systems,
representing the lowest-energy state in each spin sector. For each
spin configuration, a corresponding wave function amplitude is listed. 
A crucial detail is that all coefficients are real-valued and positive definite; 
this is a consequence of the model's frustration-free property and is what allows for
sign-problem-free MC simulation. The configurations that contribute consist of
all non-crossing, non-nested placements of $c = S^z_\text{tot} \ge 0$
up-canted bonds, with the spin background being of Dyck word form everywhere else
(as illustrated in Fig.~\ref{FIG:excited-bond-basis-states}). For the Fredkin
model, mismatch states ($m > 0$) are always at higher energy and do not play a
role in the lowest-lying excitations.
}
\begin{tabular}{
c S[table-format=1.6,group-digits=false] 
c S[table-format=1.7,group-digits=false] 
c S[table-format=1.7,group-digits=false]
c S[table-format=1.0,group-digits=false] }
\hline\hline\\[-1em]
\multicolumn{2}{c}{$S^z_\text{tot}=0$} & \multicolumn{2}{c}{$S^z_\text{tot}=1$} & \multicolumn{2}{c}{$S^z_\text{tot}=2$} & \multicolumn{2}{c}{$S^z_\text{tot}=3$}\\[0.2em] \hline
$\text{Configuration}$ & $\text{Amplitude}$ &
$\text{Configuration}$ & $\text{Amplitude}$ &
$\text{Configuration}$ & $\text{Amplitude}$ &
$\text{Configuration}$ & $\text{Amplitude}$ \\ \hline
$\po\po\po\pc\pc\pc = \up\up\up\dn\dn\dn$ & 0.447214 & $\co\cc\po\po\pc\pc = \up\up\up\up\dn\dn$ & 0.71103 & $\co\cc\co\cc\po\pc = \up\up\up\up\up\dn$ & \text{1}\\
$\po\po\pc\po\pc\pc = \up\up\dn\up\dn\dn$ & 0.447214 & $\co\cc\po\pc\po\pc = \up\up\up\dn\up\dn$ & 0.621428\\
$\po\pc\po\po\pc\pc = \up\dn\up\up\dn\dn$ & 0.447214 & $\co\po\pc\cc\po\pc = \up\up\dn\up\up\dn$ & 0.2856\\
$\po\po\pc\pc\po\pc = \up\up\dn\dn\up\dn$ & 0.447214 & $\po\pc\co\cc\po\pc = \up\dn\up\up\up\dn$ & 0.16339\\
$\po\pc\po\pc\po\pc	= \up\dn\up\dn\up\dn$ & 0.447214\\[0.15cm]
$\po\po\po\po\pc\pc\pc\pc = \up\up\up\up\dn\dn\dn\dn$ & 0.267261 & 
$\co\cc\po\po\po\pc\pc\pc = \up\up\up\up\up\dn\dn\dn$ & 0.427328 & $\co\cc\co\cc\po\po\pc\pc = \up\up\up\up\up\up\dn\dn$ & 0.719242
& $\co\cc\co\cc\co\cc\po\pc = \up\up\up\up\up\up\up\dn$ & \text{1}\\
$\po\po\po\pc\po\pc\pc\pc = \up\up\up\dn\up\dn\dn\dn$ & 0.267261 & 
$\co\cc\po\po\pc\po\pc\pc = \up\up\up\up\dn\up\dn\dn$ & 0.407145 & $\co\cc\co\cc\po\pc\po\pc = \up\up\up\up\up\dn\up\dn$ & 0.624818\\
$\po\po\pc\po\po\pc\pc\pc = \up\up\dn\up\up\dn\dn\dn$ & 0.267261 & 
$\co\cc\po\po\pc\pc\po\pc = \up\up\up\up\dn\dn\up\dn$ & 0.380996 & $\co\cc\co\po\pc\cc\po\pc = \up\up\up\up\dn\up\up\dn$ & 0.271914\\
$\po\pc\po\po\po\pc\pc\pc = \up\dn\up\up\up\dn\dn\dn$ & 0.267261 & 
$\co\cc\po\pc\po\po\pc\pc = \up\up\up\dn\up\up\dn\dn$ & 0.354469 & $\co\cc\po\pc\co\cc\po\pc = \up\up\up\dn\up\up\up\dn$ & 0.119529\\
$\po\po\po\pc\pc\po\pc\pc = \up\up\up\dn\dn\up\dn\dn$ & 0.267261 & 
$\co\cc\po\pc\po\pc\po\pc = \up\up\up\dn\up\dn\up\dn$ & 0.349927 & $\co\po\pc\cc\co\cc\po\pc = \up\up\dn\up\up\up\up\dn$ & 0.0552897\\
$\po\po\pc\po\pc\po\pc\pc = \up\up\dn\up\dn\up\dn\dn$ & 0.267261 & 
$\co\po\pc\cc\po\po\pc\pc = \up\up\dn\up\up\up\dn\dn$ & 0.277393 & $\po\pc\co\cc\co\cc\po\pc =  \up\dn\up\up\up\up\up\dn$ & 0.0318226\\
$\po\pc\po\po\pc\po\pc\pc = \up\dn\up\up\dn\up\dn\dn$ & 0.267261 & 
$\co\po\pc\cc\po\pc\po\pc = \up\up\dn\up\up\dn\up\dn$ & 0.245651\\
$\po\po\pc\pc\po\po\pc\pc = \up\up\dn\dn\up\up\dn\dn$ & 0.267261 & 
$\po\pc\co\cc\po\po\pc\pc = \up\dn\up\up\up\up\dn\dn$ & 0.237598\\
$\po\pc\po\pc\po\po\pc\pc = \up\dn\up\dn\up\up\dn\dn$ & 0.267261 & 
$\po\pc\co\cc\po\pc\po\pc = \up\dn\up\up\up\dn\up\dn$ & 0.206479\\
$\po\po\po\pc\pc\pc\po\pc = \up\up\up\dn\dn\dn\up\dn$ & 0.267261 & 
$\co\po\pc\po\pc\cc\po\pc = \up\up\dn\up\dn\up\up\dn$ & 0.0938589\\
$\po\po\pc\po\pc\pc\po\pc = \up\up\dn\up\dn\dn\up\dn$ & 0.267261 & 
$\po\pc\co\po\pc\cc\po\pc = \up\dn\up\up\dn\up\up\dn$ & 0.0855637\\
$\po\pc\po\po\pc\pc\po\pc = \up\dn\up\up\dn\dn\up\dn$ & 0.267261 & 
$\co\po\po\pc\pc\cc\po\pc = \up\up\up\dn\dn\up\up\dn$ & 0.0646068\\
$\po\po\pc\pc\po\pc\po\pc = \up\up\dn\dn\up\dn\up\dn$ & 0.267261 & 
$\po\pc\po\pc\co\cc\po\pc = \up\dn\up\dn\up\up\up\dn$ & 0.0338349\\
$\po\pc\po\pc\po\pc\po\pc = \up\dn\up\dn\up\dn\up\dn$ & 0.267261 & 
$\po\po\pc\pc\co\cc\po\pc = \up\up\dn\dn\up\up\up\dn$ & 0.0232899\\
\hline\hline
\end{tabular}
\end{table*}

Our formulation of the MC sampling algorithm relies on a representation of the
ground state in terms of noncrossing bonds that link the spins in pairs
(spin-up and spin-down, reading left to right). See
Fig.~\ref{FIG:dyck-word}(a). These bonds do not represent entangled pairs, as
in the singlet-product basis commonly used for SU(2)
systems~\cite{Beach-NPB-06, Alet-PRL-07, Beach-PRL-07, Beach-PRB-09a,
Beach-PRB-09b, Albuquerque-PRB-11, Kallin-PRB-11, Zhang-PRB-13}, but there are
important formal similarities between the valence bond description and the
representation used here. (See Sect.~III of
Ref.~\onlinecite{Adhikari-PRB-20} for more on the Hilbert space and on the
relationship between the bond, matched-delimiter, and height representations.)

In the Fredkin case, where the spin symmetry is U(1) rather than SU(2), the
ground state has strong ferromagnetic correlations and can be understood (at
least in the bulk) as a coherent spin rotor fluctuating in the $xy$ spin
plane. In the lowest-lying excited state, one spin pair is promoted (via
single spin flip) to an excited bond that carries a net positive spin
projection, representing a canting out of the $xy$ spin plane. See
Fig.~\ref{FIG:dyck-word}(b).

For a bonded pair with end points at $i < j$, the promotion of a conventional
planar bond $\po\pc=\,\uparrow_i\downarrow_j\,= \up_i\dn_j$ to an excited
canted bond $\co\cc=\,\uparrow_i\uparrow_j\,= \up_i\up_j$ is effected by
$\hat{S}^+_j$, and it is tempting to view the excitation as localized at $j$,
the position of the spin flip. Instead, we emphasize that the excited bond is
a partially delocalized object with nontrivial spatial extent $\langle j-i
\rangle$. The evolution of the positions $i$ and $j$ are correlated with each
other and with the background sea of conventional bonds in which they move.
According to the rules of the Hamiltonian, multiple such bonds cannot move
past one another: they are order-preserving and perform so-called single-file
dynamics.

We show that it is also possible to construct a mean-field version of this
theory using an effective-medium picture. At this level of approximation, it
is straightforward to solve for the wave function of the single excited bond
and to understand its transport as essentially diffusive in (imaginary) MC
time, but with an effective diffusion constant that is nontrivially enhanced
because of its motion through the background soup of conventional bonds (a
process that is subject to a combination of local and global constraints). The
computed mean-field value $z_{\text{mf}} \doteq 2.52(1)$ provides a lower
bound on the true dynamical exponent. The mean-field wave function also serves
as an excellent starting point for the projective MC simulation. Our analysis
of the right-edge first passage times of the $\sztot=1$ excited bond produces
a numerical estimate of $z \doteq 3.16(1)$.

The paper is structured as follows. Section~\ref{SECT:Model} reviews the key
details of the Fredkin model and of the pair-product basis we use to span the
Hilbert space. The mean-field treatment is presented in
Sect.~\ref{SECT:Mean-field}. The projective MC sampling algorithm that
simulates the quantum dynamics of a single excitation is described in
Sect.~\ref{SECT:Monte-Carlo}. Results from various MC simulations are
collected in Sect.~\ref{SECT:Results}. Finally, Sect.~\ref{SECT:Conclusions}
offers a recapitulation and some concluding remarks.

\section{Model and Hilbert space}
\label{SECT:Model}

The Fredkin model describes an open chain of $S=1/2$ spins interacting via
three-site interactions of the form \begin{equation}
\label{EQ:Fredkin-interaction} F_i = U_{i-1}P_{i,i+1} + P_{i-1,i}D_{i+1}.
\end{equation} Here, $U_i= \ket{\uparrow_i}\bra{\uparrow_i} =
\ket{\up_i}\bra{\up_i}$ and $D_i= \ket{\downarrow_i}\bra{\downarrow_i} =
\ket{\dn_i}\bra{\dn_i}$ filter states of up and down character at site $i$,
and $P_{i,i+1}$ is the spin-singlet projector acting on adjacent sites $i$ and
$i+1$ (following the notation of Ref.~\onlinecite{Adhikari-PRB-19}). The
presence of $U_{i-1}$ and $D_{i+1}$ in Eq.~\eqref{EQ:Fredkin-interaction}
makes the action of the projector contingent on the presence of a spin up on
the left or a spin down on the right.

Opposing magnetic fields are applied at the two ends of the chain. Their
purpose is to impose a spin twist across the system and thus to prevent the
formation of a uniform, $z$-directed, ferromagnetic ground state. Instead, the
fields stabilize a state in which the spins in the bulk flop into the $xy$
plane. We focus on the particular case of infinitely strong applied fields, a
limit in which the edge spins no longer fluctuate. Hence, for a chain of even
length $N$, there are $N-2$ live spins in the chain interior (sites $i = 2, 3,
\ldots N-1$) and two dead spins at the chain edges (sites $i=1$ and $i=N$).
The former evolve under the action of the quantum Hamiltonian; the latter are
held fixed with configurations $\ket{\uparrow}_1 = \ket{\up}_1$ and
$\ket{\downarrow}_N = \ket{\dn}_N$.

The Hamiltonian in this limit is
\begin{equation}
H = \sum_{i=2}^{N-1}F_i.
\end{equation}
with $F_i$ defined as per Eq.~\eqref{EQ:Fredkin-interaction}, except at the
edges, where the operators take the form
\begin{equation}
\begin{split}
F_2 &= P_{2,3} + \frac{1}{2}D_2D_3,\\
F_{N-1} &= P_{N-2,N-1} + \frac{1}{2}U_{N-2}U_{N-1}.
\end{split}
\end{equation}
The Fredkin Hamiltonian is engineered to produce a zero-energy ground state,
whose wave function is an equal-weight superposition of spin configurations
that form a balanced string:
\begin{equation} \label{EQ:Fredkin-ground-state}
\ket{\text{GS}} 
= \frac{1}{\sqrt{\mathcal{N}^{(0,0)}}}\sum_\mathcal{D} \ket{\mathcal{D}}.
\end{equation}
The wave function normalization is
$\mathcal{N}^{(0,0)} = C_{N/2}$, where $C_n$ is the Catalan number,
\[ C_n = \frac{1}{n+1}{2n \choose n}
= \frac{(2n)!}{(n+1)!n!} = \prod_{k=2}^n \frac{n+k}{k}. \]
[The superscripts adorning $\mathcal{N}^{(0,0)}$ anticipate notation that will
be introduced before Eqs.~\eqref{EQ:hilbert-space-decomposition} and
\eqref{EQ:hilbert-space-counting}.] In these so-called Dyck word states,
labeled $\mathcal{D}$ in Eq.~\eqref{EQ:Fredkin-ground-state}, the up and down
spins are matched and properly nested. For example, the first column of
Table~\ref{TABLE:low-energy-states-N6-N8} shows the 5~(14) relevant states for
$N = 6$~($N=8$).

\begin{table}
\caption{\label{TABLE:all-delimiters-N8} 
Here, all $\sztot \ge 0$ states of the 8-site Fredkin model's Hilbert space
are expressed in the delimiter notation. The entries correspond to the first
four rows of Fig.~2 in Ref.~\onlinecite{Adhikari-PRB-20} (although differently
ordered). For a given value of the spin projection, the state of lowest energy lives within the
no-mismatch ($m=0$) sector.
}
\begin{tabular}{ccc}
\hline\hline\\[-0.31cm]
{$\sztot$} & {$m$} & Basis elements \\[0.05cm] \hline
3 & 0 & $\co\cc\co\cc\co\cc\po\pc$\\[0.15cm] 
\multirow[t]{2}{*}{2} & \multirow[t]{2}{*}{0} & $\co\cc\co\cc\po\po\pc\pc$,\ $\co\cc\co\cc\po\pc\po\pc$,\ $\co\cc\co\po\pc\cc\po\pc$,\\ 
  &   & $\co\cc\po\pc\co\cc\po\pc$,\ $\co\po\pc\cc\co\cc\po\pc$,\ $\po\pc\co\cc\co\cc\po\pc$ \\[0.15cm] 
\multirow[t]{4}{*}{1} & 1 & $\po\pc\mo\mc\co\cc\po\pc$\\[0.15cm] 
  & \multirow[t]{3}{*}{0} & $\po\pc\po\pc\co\cc\po\pc$,\ $\po\pc\co\po\pc\cc\po\pc$,\ $\po\pc\co\cc\po\pc\po\pc$,\ 
        $\po\pc\co\cc\po\po\pc\pc$,\ $\po\po\pc\pc\co\cc\po\pc$,\\ 
  &   & $\co\po\pc\po\pc\cc\po\pc$,\ $\co\po\pc\cc\po\pc\po\pc$, $\co\cc\po\po\po\pc\pc\pc$,\ 
        $\co\cc\po\po\pc\po\pc\pc$,\ $\co\cc\po\po\pc\pc\po\pc$,\\ 
  &   & $\co\cc\po\pc\po\po\pc\pc$,\ $\co\cc\po\pc\po\pc\po\pc$,\ $\co\po\po\pc\pc\cc\po\pc$,\ $\co\po\pc\cc\po\po\pc\pc$ \\[0.15cm] 
\multirow[t]{5}{*}{0} & 2 & $\po\pc\mo\mo\mc\mc\po\pc$ \\[0.15cm] 
  & 1 & $\po\pc\mo\po\pc\mc\po\pc$,\ $\po\pc\mo\mc\po\pc\po\pc$,\ $\po\pc\mo\mc\po\po\pc\pc$,\ $\po\pc\po\pc\mo\mc\po\pc$,\ $\po\po\pc\pc\mo\mc\po\pc$ \\[0.15cm] 
  & \multirow[t]{3}{*}{0} & $\po\pc\po\pc\po\pc\po\pc$,\ $\po\pc\po\pc\po\po\pc\pc$,\ $\po\pc\po\po\pc\pc\po\pc$,\ 
        $\po\pc\po\po\pc\po\pc\pc$,\ $\po\pc\po\po\po\pc\pc\pc$,\\  
  &   & $\po\po\pc\pc\po\pc\po\pc$,\ $\po\po\pc\pc\po\po\pc\pc$,\ $\po\po\pc\po\pc\pc\po\pc$,\ $\po\po\pc\po\pc\po\pc\pc$,\ $\po\po\pc\po\po\pc\pc\pc$, \\ 
  &   & $\po\po\po\pc\pc\pc\po\pc$,\ $\po\po\po\pc\pc\po\pc\pc$,\ $\po\po\po\pc\po\pc\pc\pc$,\ $\po\po\po\po\pc\pc\pc\pc$\\
\hline\hline
\end{tabular}
\end{table}

In terms of the height function $h_i = \sum_{\!j=1}^{i} \sigma^z_j$, the
configurations contributing to the ground state satisfy $h_0 = h_N = 0$ and
$h_1 = h_{N-1} = 1$ (see Fig.~\ref{FIG:dyck-word}), and the height function is
everywhere non-negative. That is to say, the configurations correspond to
landscapes that begin and end at the horizon and never drop below it. Excited
states differ from this prescription in two ways: they may have a nonzero
$z$-directed net magnetization, in which case the height profile does not
return to the horizon ($h_N \neq 0$); or they may break the Dyck word
property, with the violation corresponding to a valley region where the height
profile goes negative.

The Hamiltonian has several useful symmetry properties. It commutes with
$S^z_\text{tot} = \sum_i S^z_i$, so the Hamiltonian is block diagonal in a
basis of states of definite total $z$ spin projection. More important, the
Hamiltonian also preserves the number of pair-bond mismatches in the Dyck word
sense. As a consequence, the Hilbert space can also be made to break into
distinct sectors based on the decomposition of each spin configuration into a
product of pair bonds.

In order to span the full Hilbert space, four bond species are required. We
denote them by parentheses, angled brackets, and the ceiling and floor
brackets:
\begin{equation}
\po\pc = \up\dn,\
\mo\mc = \dn\up,\
\co\cc = \up\up,\
\aco\acc = \dn\dn.
\end{equation}
The four bond types shown correspond to the conventional $xy$ {\it planar}
bond (Dyck-word compatible), the {\it mismatch} bond (a Dyck-word defect), and
the two {\it canted} bonds that tilt up and down out of the $xy$ plane. In our
convention, the parentheses are matched and properly nested but otherwise
unrestricted; the angled brackets are strictly Matryoshka nested; and the
ceiling and floor brackets (either $\co\cc$ or $\aco\acc$ appear, but not
both) are matched but never nested. These rules produce the correct state
counting. Because of the hard boundary conditions, the right end point of an
up-canted bond $\cc\!=\!\up$ can never occupy the rightmost site of the chain
segment, and the left end point of a down-canted bond $\aco\!=\!\dn$ can never
occupy the leftmost site. Similar considerations demand that mismatch bonds
always be located in the chain interior, never on the boundary.

As emphasized elsewhere [see, e.g., Eq.~(12) of
Ref.~\onlinecite{Caha-arXiv-18}], the local Fredkin interaction $\mathsf{s}_j
= (\mathbb{1} - 2F_j)$ acts as a short-bond-shuffle operation. The only
allowed rearrangements are the following:
\begin{alignat}{3} \label{EQ:short-bond-shuffle}
\nonumber
\mathsf{s}\ket{\up\up\dn} &=\ket{\up\dn\up} & \ \ \iff \ \ \mathsf{s}\ket{\po\pc x} &= \ket{x \po\pc}\\ 
\nonumber
\mathsf{s}\ket{\up\dn\up} &= \ket{\up\up\dn} & \mathsf{s}\ket{x \po\pc} &= \ket{\po\pc x}\\
\mathsf{s}\ket{\up\dn\dn} &= \ket{\dn\up\dn}\\
\nonumber
\mathsf{s}\ket{\dn\up\dn} &= \ket{\up\dn\dn}
\end{alignat}
Here, $x$ is any spin. The other four possible configurations
($\ket{\up\up\up}, \ket{\dn\up\up}, \ket{\dn\dn\up}, \ket{\dn\dn\dn}$) are
left unchanged. Hence, under action of the Hamiltonian, all states can be
organized into equivalence classes of the form
\begin{subequations} 
\begin{equation} \label{EQ:equiv-class-up}
\underbrace{\!\po\pc\po\pc\cdots\po\pc\!}_{2p-2}\,\,\underbrace{\!\mo\mo\cdots\mc\mc\!}_{2m}\,\,\underbrace{\!\co\cc\co\cc\cdots\co\cc\!}_{2c}\,\po\pc
\end{equation}
or
\begin{equation} \label{EQ:equiv-class-dn}
\po\pc\,\underbrace{\!\aco\acc\aco\acc\cdots\aco\acc\!}_{2c}
\,\,\underbrace{\!\mo\mo\cdots\mc\mc\!}_{2m}
\,\,\underbrace{\!\po\pc\po\pc\cdots\po\pc\!}_{2p-2}
\end{equation}
\end{subequations}
corresponding to the $S^z_\text{tot} \ge 0$ and $S^z_\text{tot} < 0$ cases,
respectively. In each of Eqs.~\eqref{EQ:equiv-class-up} and
\eqref{EQ:equiv-class-dn}, there are $N$ delimiter symbols in total, one
assigned to each site of the spin chain: these comprise $2p$ matched
parenthesis ($p-1$ pairs arranged on one side and $1$ pair on the other), $2m$
nested angled brackets, and $2c$ matched square brackets (either all ceiling
or all floor). Expressed in the landscape language, the number of mismatches
is connected to the lowest level achieved below the horizon via $m =
\min(0,h_N)-\min\{h_i\}$; the elevation at the right edge is connected to the
spin sector through $c = \lvert h_N \rvert/2 = \lvert\sztot\rvert$. A more
detailed discussion of these relations appears in Sect.~III of
Ref.~\onlinecite{Adhikari-PRB-20}. Table~\ref{TABLE:all-delimiters-N8} shows
the various states in delimiter notation for the $N=8$ system.

For a chain of even length $N$, there are $p$ planar bonds, $m$ mismatches,
and $c$ canted bonds, subject to the tiling constraint $p+m+c=N/2$. This is
slightly different from the counting in Eq.~(7) of
Ref.~\onlinecite{Salberger-arXiv-16}, since we are working in the limit of
infinitely strong external fields where the first and last spin in the chain
are fixed. In the no-mismatch case $(m=0$), the number of planar bonds ranges
over $p = 1, 2, \ldots, N/2$ and all spin projection values short of full
polarization are achievable: $\lvert\sztot\rvert = c = N/2-p = 0, 1, \ldots
N/2-1$. On the other hand, for $1 \le m \le N/2-2$, the number of planar bonds
ranges over $p = 2, 3, \ldots, N/2-m$, and the maximum spin polarization is
reduced for each mismatch bond present: $\lvert\sztot\rvert = c = N/2-m-p = 0,
1, \ldots N/2-1-m$.

All of this is to say that the Hilbert space can be decomposed into disjoint
sectors, according to
\begin{equation} \label{EQ:hilbert-space-decomposition}
\begin{split}
&\mathcal{H} = \bigoplus_{m,c} \mathcal{H}^{(m,c)} = \mathcal{H}^{(0,0)}
\oplus \mathcal{H}^{(0,1)}
\oplus \mathcal{H}^{(0,1)}
\oplus  \cdots,\\
&2^{N-2} 
= \dim{\mathcal{H}}
= \mathcal{N}^{(0,0)} + 2\sum_{m \ge 0} \sum_{c \ge 1} \mathcal{N}^{(m,c)}.
\end{split}
\end{equation}
In this accounting, each block with $c \neq 0$ must appear twice, with one
block representing the up-canted states and one the down-canted. $\mathcal{N}^{(m,c)} =
\dim\mathcal{H}^{(m,c)}$ is the number of valid configurations with $m$
mismatches and $c$ canted bonds. The basis size within the various bond
sectors is as follows:
\begin{equation}
\label{EQ:hilbert-space-counting}
\begin{split}
\mathcal{N}^{(0,0)} &= \frac{1}{N/2+1}{N \choose N/2} = C_\frac{N}{2},\\
\mathcal{N}^{(0,1)} &= \frac{4}{N/2+2}{N - 1\choose N/2-2} = C_{\frac{N}{2}+1} - 2C_{\frac{N}{2}},\\
\mathcal{N}^{(1,0)} &= \frac{5}{N/2+2}{N-2 \choose N/2-3},\\
\mathcal{N}^{(1,1)} &= \frac{7}{N/2+3}{N-2 \choose N/2-4},\ \ldots
\end{split}
\end{equation}
In particular, $\mathcal{N}^{(0,0)}$ and $\mathcal{N}^{(0,1)}$ represent the
ground-state and lowest-lying excited-state sectors of the Fredkin model. We
find that they have comparable scaling, both growing exponentially in the
system size. The sector with a single canted bond has roughly double the
number of basis states as the ground state: $\mathcal{N}^{(0,1)} /
\mathcal{N}^{(0,0)} = 2 - 12/N + 48/N^2 + O(N^{-3})$, since
\[ \mathcal{N}^{(0,0)} = C_{N/2} = 2^{N+1}\biggl(\frac{\sqrt{\,2/\pi\,}}{(3/2+N)^{3/2}} + O(N^{-7/2})\biggr)\]
and
\begin{equation} \label{EQ:size-N01}
\begin{aligned}
\mathcal{N}^{(0,1)} 
&=\sum_{\lceil i,j \rceil} C_{\frac{i-1}{2}} C_{\frac{j-i-1}{2}} C_{\frac{N-j}{2}}
=C_{\frac{N}{2}+1} - 2C_{\frac{N}{2}}\\
&= 2^{N+2}\biggl(\frac{\sqrt{\,2/\pi\,}}{(11/2+N)^{3/2}}+ O(N^{-7/2})\biggr).
\end{aligned}
\end{equation}
In Eq.~\eqref{EQ:size-N01}, the sum $\lceil i,j \rceil$ 
ranges over the $N(N-2)/8$ allowed canted bond positions, i.e.,
$i=1,3,\ldots,N-3$ and $j=i+1,i+3,\ldots N-2$.

The ground state of the Fredkin spin chain is unique. It has spin projection
$\sztot=0$ and energy $E_0 = 0$. The first excited state is doubly degenerate,
with $\sztot = \pm 1$; all contributing spin configurations are smoothly
connected to either
$\po\pc\cdots\po\pc\co\cc\po\pc$
or
$\po\pc\aco\acc\po\pc\cdots\po\pc$.
Its energy is $E_1 = E_0 + \Delta = O(N^{-z})$. The dynamical exponent has
been estimated, both analytically and with DMRG~\cite{Adhikari-PRB-19, Bravyi-PRL-12, DellAnna-PRB-16, Chen-PRB-17,
Chen-JPA-17}, and the most recent values are in the range $3.0 \lesssim z \lesssim 3.2
$~\cite{Adhikari-PRB-19}. Chen and coworkers have offered arguments for the
Fredkin model's slow dynamics based on an analogy with classical spin
chains~\cite{Chen-PRB-17}. Nonetheless, a full understanding of the physical
mechanism responsible for the enhancement of $z$ has not yet been achieved.

\section{Mean-field treatment}
\label{SECT:Mean-field}

\begin{figure}
\begin{center}
\includegraphics[width = 0.9\columnwidth]{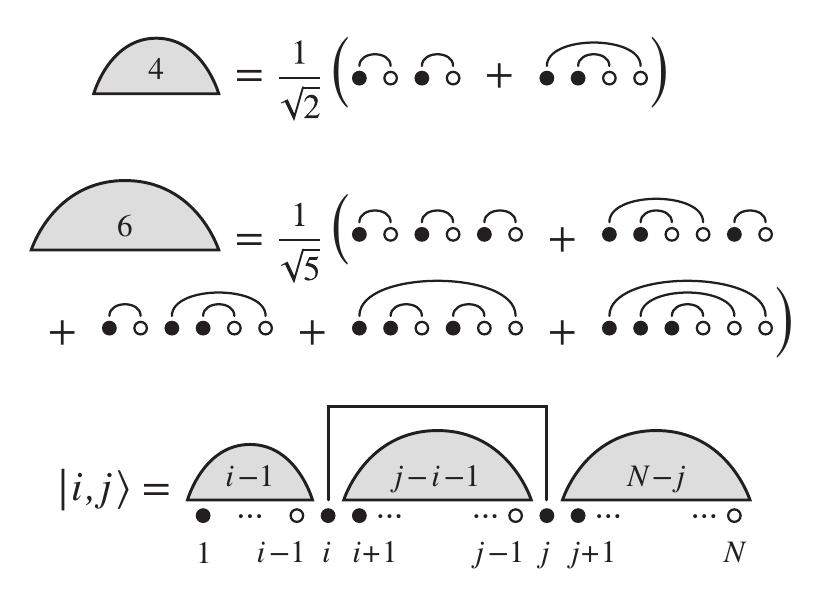}
\end{center}
\caption{\label{FIG:excited-bond-basis-states} 
The states $\{\ket{i,j} : \text{$i$ odd; $j$ even; $i < j < N-1$}\}$, defined
in Eq.~\eqref{EQ:Excited-bond}, serve as an approximate basis for describing
the low-lying states in the $S^z_\text{tot}=1$ spin sector.
}
\end{figure}

The ground-state wave function of the Fredkin model is an equal-weight
superposition of the Dyck word configurations, numbering
$\mathcal{N}^{(0,0)}$, that constitute the $m=c=0$ sector of the Hilbert
space. This is expressed in Eq.~\eqref{EQ:Fredkin-ground-state}. The states
next lowest in energy are $\sztot = \pm 1$ excitations that reside in the
$m=0$, $c=1$ sector.
 
While, in general, one can achieve $\sztot = \pm 1$ by flipping a single spin
in any $\sztot=0$ configuration, not all such transformations produce
configurations that contribute to the low-lying excited state. Specifically,
an injudicious choice may also generate an unwanted
mismatch bond. To target the desired sector, the correct procedure
is to  (i) select a Dyck word state, (ii)
identify a bond that is not enclosed by any other bond, and (iii) promote that
bond according to $\po\cdots\pc \to \co\cdots\cc$ (or $\aco\cdots\acc$). For
$\sztot=-1$ ($+1$), an equivalent prescription is to select a site $i$ ($j$)
away from the chain edges that is ascending from (descending to) the landscape horizon, i.e.,
$h_i = 1$ for $i \in \{3,5,\ldots,N-1\}$ ($h_j=0$ for $j \in \{2,4,\ldots,N-2\}$), and to flip the spin there from up (down) to down (up); this is guaranteed to produce a zero-mismatch state.




The states produced in this way span the Hilbert subspace $\mathcal{H}^{(0,1)}$. In the up-canted
case, they have spin configurations of the form
$(1, \sigma^z_2, \ldots, \sigma^z_{i-1},
1, \sigma^z_{i+1}, \ldots,  \sigma^z_{j-1},\allowbreak
1, \sigma^z_{j+1}, \ldots, \sigma^z_{N-1}, -1)$,
such that each of the subsequences
$\mathcal{D}' = (\sigma^z_1, \ldots, \sigma^z_{i-1})$,
$\mathcal{D}'' = (\sigma^z_{i+1}, \ldots,  \sigma^z_{j-1})$,
and
$\mathcal{D}''' = (\sigma^z_{j+1}, \ldots, \sigma^z_{N})$
is itself a Dyck word. Alternatively, we could refer to the set of height profiles with
$h_0 = h_{i-1}=0$,
$h_1 = h_i = 1$,
$h_j = h_{N} = 2$,
and $h_k \ge 0$ otherwise.

With this in mind, we introduce a useful mean-field ansatz. We start from the
most general form of the excited state. We then simplify it by focusing on
the location of the canted bond and averaging over the background spin
texture. This is equivalent to the construction shown in
Fig.~\ref{FIG:excited-bond-basis-states} in which a Fredkin ground state of
the corresponding finite chain segment is inserted everywhere there is not a
canted bond:
\begin{equation}\label{EQ:Excited-state}
\begin{aligned}
\lvert \text{ES} \rangle 
&= \sum_{\sigma^z \in \mathcal{H}^{(0,1)}} g(\sigma^z_1,\sigma^z_2,\ldots,\sigma^z_N) \ket{\sigma^z_1\sigma^z_2\cdots \sigma^z_N}
\\[-0.65em]
&=
\sum_{\lceil i,j \rceil,\{\mathcal{D}\}} g_{ij}(\{\mathcal{D}\}) 
\ket{\,\overbrace{\underbrace{\!\!\po \cdots \pc\!\!}_{i-1}}^{\mathcal{D}'}\,\co_i\,
\overbrace{\underbrace{\!\!\po \cdots \pc\!\!}_{j-i-1}}^{\mathcal{D}''}\,\cc_j\,
\overbrace{\underbrace{\!\!\po \cdots \pc\!\!}_{N-j}}^{\mathcal{D}'''}\,}\\
&\approx
\sum_{\lceil i,j \rceil} \bar{g}_{ij} \underbrace{\sum_{\{\mathcal{D}\}} \ket{\,\underbrace{\!\!\po \cdots \pc\!\!}_{\mathcal{D}'}\,\co_i\,
\underbrace{\!\!\po \cdots \pc\!\!}_{\mathcal{D}''}\,\cc_j\,
\underbrace{\!\!\po \cdots \pc\!\!}_{\mathcal{D}'''}\,}}_{\ket{i,j}}\\[-0.65em]
&\sim \sum_{i,j}\bar{g}_{ij}\ket{i,j}.
\end{aligned}
\end{equation}
In the second line of Eq.~\eqref{EQ:Excited-state}, the Dyck word
$\mathcal{D}'$ is a spin segment of length $i-1$; similarly, $\mathcal{D}''$
and $\mathcal{D}'''$ are segments of length $j-i-1$ and $N-j$. The sum over
all canted bond placements includes the possibilities that $i=1$ or $j=i+1$,
implying that $\mathcal{D}'$ or $\mathcal{D}''$ may be of length zero. In the
approximation on the third line, $\bar{g}_{ij}$ is the mean-field wave
function that gives the probability amplitude to find a canted bond with end
points at the given indices. We are assuming that the general amplitude
$g_{ij}(\{\mathcal{D}\}) = g_{ij}(\mathcal{D}',\mathcal{D}'',\mathcal{D}''') \approx \bar{g}_{ij}$
depends only on the excited bond position and is largely independent of the
nature of the Dyck word segments that surround sites $i$ and $j$.

One can view this approach as relying on a separation of time scales.
The assumption is that the churn of the background of planar bonds occurs on a
much faster time scale than the motion of the canted bond, so that when the
canted bond moves ($i \to i \pm 2$ or $j \to j \pm 2$) the background quickly
relaxes to a pure Fredkin ground state in each of the disjoint spin chain segments. This says,
in essence, that the canted bond leaves no froth in its wake as it travels. In practice, we
are ignoring entropic contributions from the disruption to the background, and
so we are overestimating the excitation's mobility and underestimating the
value of the dynamical exponent.

The mean-field approximation to the excited state is given by $\lvert
\psi_\text{mf} \rangle = \sum_{i,j}\bar{g}_{ij}\ket{i,j}$. The mean-field
basis states
\begin{multline}\label{EQ:Excited-bond}
\ket{i,j} = \frac{1}{\sqrt{C_{\frac{i-1}{2}}C_{\frac{j-i-1}{2}}C_{\frac{N-j}{2}}}}
\sum_{\{\mathcal{D}\}}\ket{\mathcal{D}_{1,i-1}} \otimes \ket{\uparrow_i}\\
\otimes \ket{\mathcal{D}^{\prime \prime}_{i+1,j-1}} \otimes \ket{\uparrow_j} \otimes \ket{\mathcal{D}^{\prime \prime \prime}_{j+1,N}}
\end{multline}
are $N(N-2)/8$ in number and are defined so as to be orthonormal,
$\overlap{k,l}{i,j} = \delta_{i,k}\delta_{j,l}$. This basis retains much of
the expressive power of the full $m=0$, $c=1$ basis but is radically smaller:
$O(N^2) \ll \mathcal{N}^{(0,1)} = O(2^N/N^{3/2})$. A consequence is that the
Lanczos method can be used to determine the mean-field ground state in as few
as $O(N^4)$ operations; by contrast, the equivalent computation in the full
basis cannot be completed in polynomial time.

For our purposes, it is convenient to define a discrete-time-step evolution
operator that implements the Fredkin Hamiltonian's short-bond-shuffle dynamics.
\begin{equation}\label{EQ:H-shift}
	\mathsf{U} = (N-2)\mathbb{1} - 2H = \sum_{j=2}^{N-1}\Bigl(\mathbb{1}-2F_j\Bigr) = \sum_{j=2}^{N-1}\mathsf{s}_j,
\end{equation}
with the diagonal $N-2$ contribution providing the minimum shift necessary to
make all the matrix elements positive definite. Letting $\mathsf{U}$ act on
$\ket{i,j}$ either leaves the state as is or produces a 
superposition of the states $\ket{i\pm 2,j}$
and $\ket{i,j\pm 2}$. The end points of the canted bond move by two sites
whenever they are able to shuffle past a short bond, i.e., $\co \po \pc
\leftrightarrow \po \pc \co$ or $\cc \po \pc \leftrightarrow \po \pc \cc$,
according to Eq.~\eqref{EQ:short-bond-shuffle}

It is entirely practical to solve the
matrix eigenproblem
$\mathsf{U}\bar{g} = (N-2-2E_1)\bar{g}$.
We simply need to compute the relevant matrix elements:
\begin{equation}
\begin{split}
\bra{3,j}\mathsf{U}\ket{1,j} &= \sqrt{\,\mathcal{P_\text{s}}(j-2)},\\
\bra{i\pm 2,j}\mathsf{U}\ket{i,j} &= \sqrt{\,\mathcal{P_\text{s}}(i\pm 1)\mathcal{P}_\text{s}(j-i\mp 1)},\\
\bra{i,j \pm 2}\mathsf{U}\ket{i,j} &= \sqrt{\,\mathcal{P_\text{s}}(j-i\pm 1)\mathcal{P}_\text{s}(N-j+1\mp 1)}.
\end{split}
\end{equation}
These are defined in terms of the probability of finding a short conventional bond $\po \pc$
at the edge of a Dyck word of length $n$,
\begin{equation}
\mathcal{P_\text{s}}(n) = \frac{C_{\frac{n-2}{2}}}{C_{\frac{n}{2}}} = \frac{1}{4}\biggl(\frac{n+2}{n-1}\biggr).
\end{equation}
We have to pay special attention to the diagonal contribution
from short canted bonds when they are in the bulk,
\begin{equation}
\bra{i,i+1}\mathsf{U}\ket{i,i+1} = N-2 - \frac{1}{4}\biggl(\frac{i+1}{i-2} + \frac{N-i+1}{N-i-2}\biggr),
\end{equation}
and when they sit at their extremal positions on the spin chain segment,
\begin{equation}
\begin{split}
\bra{1,2}\mathsf{U}\ket{1,2} &= \frac{4N^2 - 21N+24}{4(N-3)},\\
\bra{N\!-\!3,N\!-\!2}\mathsf{U}\ket{N\!-\!3,N\!-\!2} &= \frac{4N^2 - 33N + 62}{4(N-5)}.
\end{split}
\end{equation}

As an example, we express $\mathsf{U}$ for the $N=6$ system,
\begin{equation}
\mathsf{U}_{N=6} = \begin{pmatrix} \tfrac{7}{2} & \tfrac{1}{\sqrt{2}} & 0 \\
\tfrac{1}{\sqrt{2}} & 1 & 1 \\
0 & 1 & 1
\end{pmatrix},
\end{equation}
with rows and columns arranged according to the basis ordering $\{ \ket{1,2},\ket{1,4},\ket{3,4} \}$.
The comparable result for $N=8$ is
\begin{equation}
\mathsf{U}_{N=8} = \begin{pmatrix} \tfrac{28}{5} & \sqrt{\tfrac{2}{5}} & 0 & 0 & 0 & 0\\
\sqrt{\tfrac{2}{5}} & \tfrac{7}{2} & \tfrac{1}{2} & 1 & 0 & 0\\
0 & \tfrac{1}{2} & 4 & 0 & \tfrac{1}{\sqrt{2}} & 0\\
0 & 1 & 0 & \tfrac{9}{2} & \tfrac{1}{\sqrt{2}} & 0\\
0 & 0 & \tfrac{1}{\sqrt{2}} & \tfrac{1}{\sqrt{2}} & 2 & \tfrac{1}{\sqrt{2}}\\
0 & 0 & 0 & 0 & \tfrac{1}{\sqrt{2}} & \tfrac{9}{2}
\end{pmatrix}
\end{equation}
with basis $\{ \ket{1,2}, \ket{1,4}, \ket{1,6}, \ket{3,4}, \ket{3,6}, \ket{5,6}  \}$.

\begin{figure}
\begin{center}
    \includegraphics[width=0.9\columnwidth]{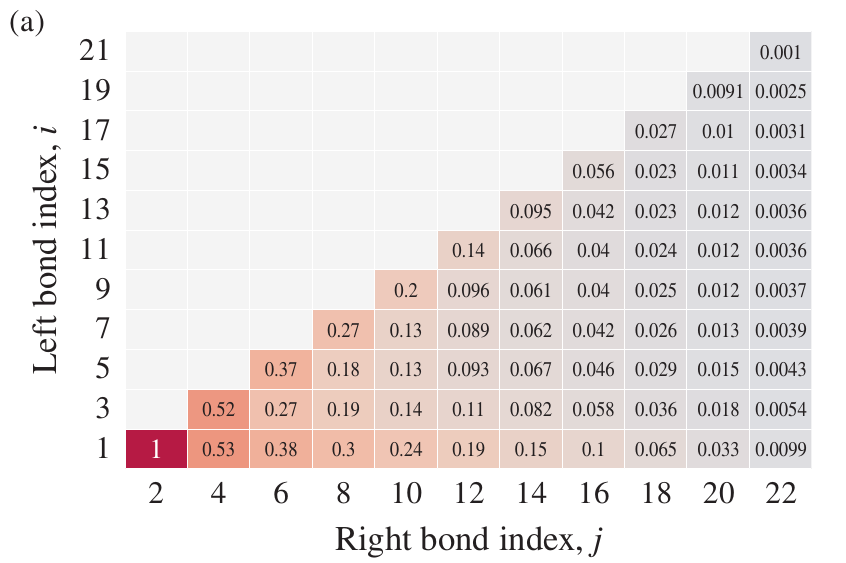}     
    \includegraphics[width=0.9\columnwidth]{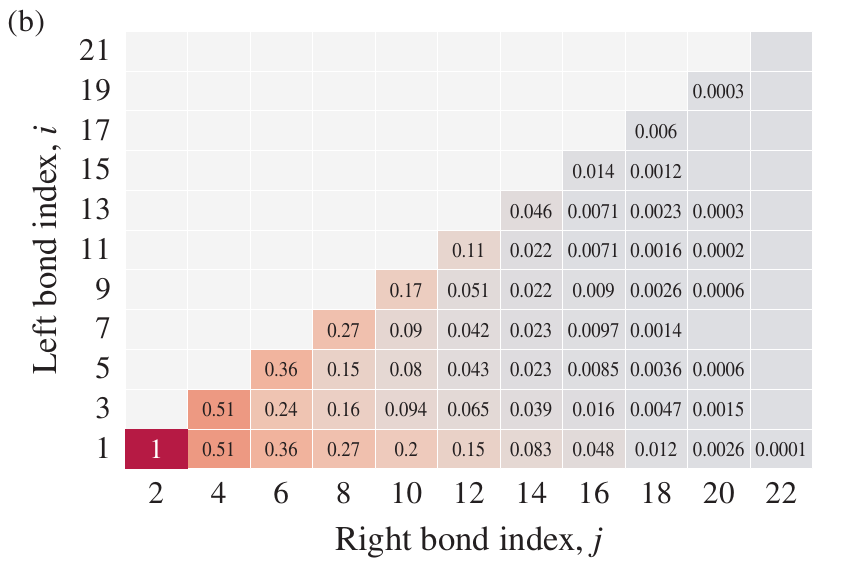}     
\end{center}   
    \caption{\label{FIG:gplot}
    The two panels show heat maps of the computed values for
    (a) the mean-field excited bond wave function $\bar{g}_{ij}$ and
    (b) the corresponding bond amplitude deduced from quantum MC simulations.
    The only meaningful entries are for $i<j$, with the excited bond 
    opening at $i$ on the left and closing at $j$ on the right.
    The data are for a Fredkin chain of length $N=24$. While not shown
    here, we have generated mean-field data for system sizes up to $N=300$.
    We find that the dominant contribution is always
    $\bar{g}_{1,2}$ and that the amplitudes fall off most slowly
    along the diagonal ($i=j-1$) and horizontal $(i=1)$ directions.
    It appears that those two directions share
    a common powerlaw decay:
    $g_{1,j} \sim g_{j-1,j} \sim (j-1)^{-2/3}$ for $2 \le j \lesssim N/3$.
    }
\end{figure}

Our numerical task is to solve the eigenequation
$\mathsf{U}\bar{g} = \lambda \bar{g}$ 
for the the largest eigenvalue $\lambda = N-2-2E_1$.
Having accounted for the transformation (shift and sign flip) of the energy eigenvalues 
implicit in Eq.~\eqref{EQ:H-shift}, we find that the energy gap is
\begin{equation}
\Delta = E_1 - E_0 = E_1 = \frac{1}{2}\bigl(N-2-\lambda\bigr).
\end{equation}
The values we obtain are shown in the inset of Fig.~\ref{FIG:survival-fraction}.
The asymptotic behavior of the gap scaling with system size
suggests $z_\text{mf} = 2.52(1)$, which we understand to be a lower
bound on the true value of the dynamical exponent.

The quality of the mean-field result can be assessed by
direct comparison to the appropriately averaged exact
wave function. Such a comparison for the 
$N=24$ system is shown in Fig.~\ref{FIG:gplot}. 
The mean field result
correctly captures the qualitative behavior,
which is that the excited bond with end points $i<j$
experiences effective attractions between the left end point of the bond and
the left edge of the chain (which favors small $i$)
and between the two bond end points themselves 
(which encourages short bonds by favoring small $j-i$).

As a practical matter, the comparison is carried out as follows.
We introduce operators $b_{[i,j]}$ 
and $b_{\{i,j\}}$ 
that detect the presence of a canted or mismatch bond 
connecting sites $i$ and $j$. 
The implementation is somewhat subtle:
in the case of $\sztot \ge 0$
(only up-canted bonds, no down-canted),
$b_{[i,j]}$ returns $1$ when 
acting on a state
in which $i,j$ are the largest
pair of indices such that
$h_i = 1$ and $h_j = 2$,
or $h_i = 3$ and $h_j = 4$,
and so on up to $h_i = 2c-1$ and $h_j = 2c$; 
it returns zero otherwise. 
On the other hand, 
$b_{\{i,j\}}$ triggers only
when $i$ and $j$ are the smallest
and largest indices, respectively,
such that $h_i = h_{j-1} = -1$,
or $h_i = h_{j-1} = -2$,
and so on up to $h_i = h_{j-1} = -m$.

\begin{figure*}[ht!]
\begin{center}
\includegraphics{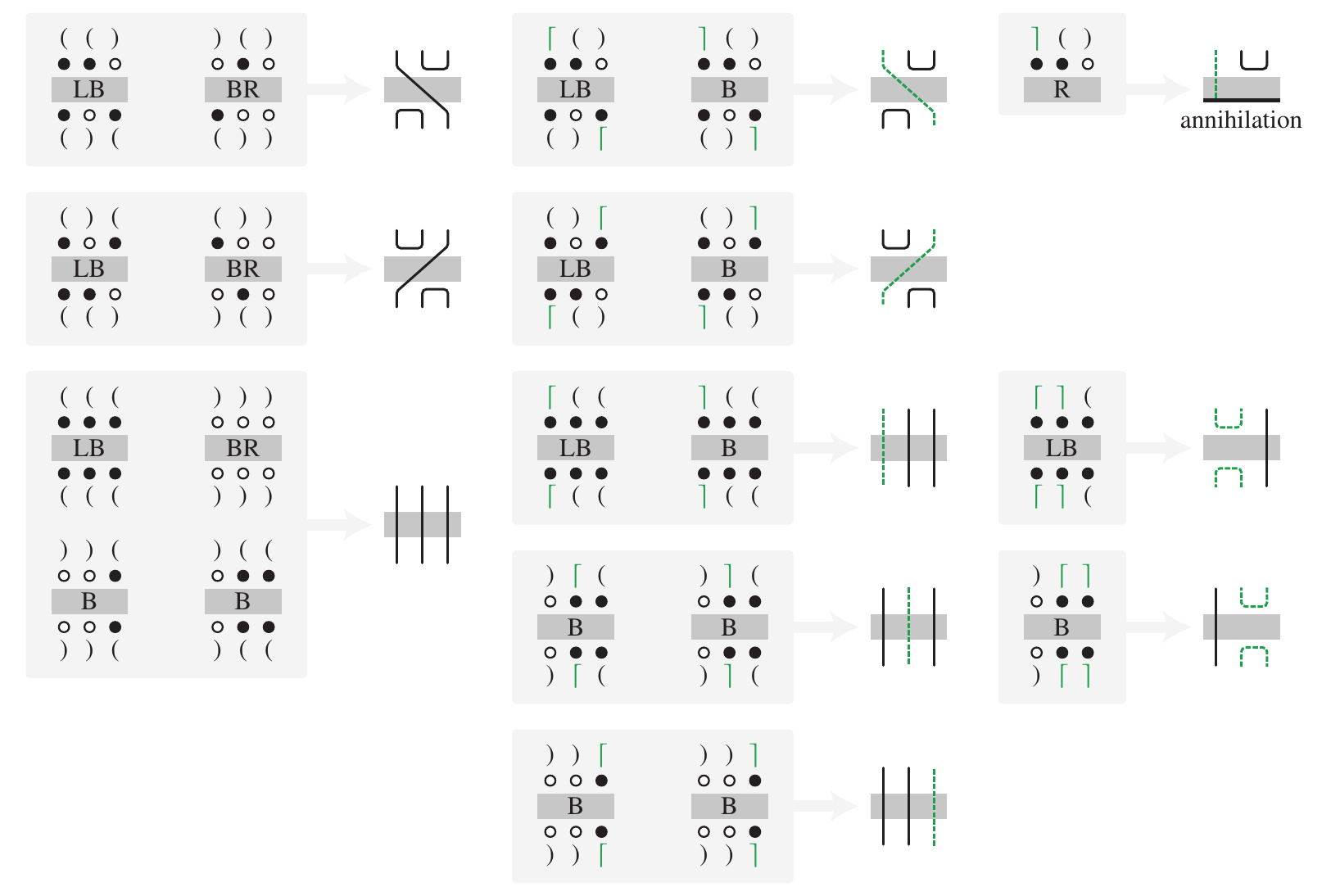}
\caption{\label{FIG:vertices}
The vertices depicted show all local rearrangements of spins
due to the operator $\mathsf{s} = (\mathbb{1} - 2F)$
for three-site groupings of sites that touch the (L) left edge of the
chain, touch the (R) right edge, or touch neither and are 
completely contained in the (B) bulk.
Light gray arrows point from a highlighted region containing multiple
vertices to a single corresponding vertex in the world-line 
representation.
In the usual way, $\up$ and $\dn$ denote up and down spins;
$\po$ and $\pc$ are the up and down end points of a conventional $xy$ planar bond;
$\co$ and $\cc$ are the both-up end points of an excited canted bond.
Black lines trace the history of planar bond end points, whereas
the
green dashed lines trace the history of canted bond end points.
}
\end{center}	 
\end{figure*}

With regard to the mean-field basis states in Eq.~\eqref{EQ:Excited-bond},
the bond detection operator satisfies
\begin{equation} 
b_{[k,l]}\ket{i,j} = \delta_{i,k}\delta_{j,l}\ket{i,j}.
\end{equation}
Hence, it has an expectation value
\begin{equation}
\frac{ 
	\bra{\psi_\text{mf}} b_{[i,j]} \ket{\psi_\text{mf}} 
}{
	\overlap{\psi_\text{mf}}{\psi_\text{mf}}
}
= \frac{\bar{g}_{i,j}^2}{\sum_{k,l} \bar{g}_{k,l}^2}.
\end{equation}
(We write $\bar{g}_{i,j}^2$ rather than $\lvert \bar{g}_{i,j} \vert^2$, since the
wave function is purely real.)
In other words, the classical probability of finding a bond
in postion $i,j$ is proportional to $\bar{g}_{i,j}^2$,
and this should be comparable to the probability to
find such a bond in the two-sided projective calculation
\begin{equation} \label{EQ:prob2-ij}
\text{prob}_2(i,j) = 
\lim_{n \gg N^z} \frac{\bra{\psi_\text{mf}}\mathsf{U}^n b_{[i,j]} \mathsf{U}^n\ket{\psi_\text{mf}}}{\bra{\psi_\text{mf}}\mathsf{U}^{2n}\ket{\psi_\text{mf}}}.
\end{equation}
For large enough values of $n$, the right-hand side of Eq.~\eqref{EQ:prob2-ij}
approaches
$\bra{\text{ES}} b_{[i,j]} \ket{\text{ES}}/
\overlap{\text{ES}}{\text{ES}}$.
Alternatively, we can compare
\begin{equation} \label{EQ:prob1-ij}
\frac{\bra{R}b_{[i,j]}\ket{\psi_\text{mf}}}{\overlap{R}{\psi_\text{mf}}}= \frac{\sqrt{C_{(i-1)/2}C_{(j-i-1)/2}C_{(N-j)/2}}\,g_{i,j}}{\sum_{k,l}\sqrt{C_{(k-1)/2}C_{(l-k-1)/2}C_{(N-l)/2}}\,g_{k,l}}
\end{equation}
to the one-sided projection
\begin{equation}
\text{prob}_1(i,j) = \lim_{n \gg N^z} \frac{\bra{R} b_{[i,j]} 
\mathsf{U}^n\ket{\psi_\text{mf}}}{\bra{R}\mathsf{U}^{n}\ket{\psi_\text{mf}}},
\end{equation}
where the bra on the left is some reference state. It is convenient to choose $\ket{R}$
to be the purely disordered paramagnetic state $\otimes_{i=1}^N\bigl(\ket{\uparrow}_i + \ket{\downarrow}_i)$, which has overlap 1 with every spin configuration.

\section{Monte Carlo Sampling}
\label{SECT:Monte-Carlo}

The lowest-energy state of the Fredkin model in each $\sztot$ sector has a wave function with real, 
positive-definite amplitudes (as shown in Table~\ref{TABLE:low-energy-states-N6-N8}). Each such state corresponds to a fixed
number of canted bonds ($c = \lvert\sztot\rvert$) and exactly zero mismatch bonds ($m=0$).
These properties allow us to formulate a 
sign-problem-free projector MC algorithm that simulates
the dynamics of one or more canted bonds
as they move through a fluctuating background of planar bonds.
We focus on the case of a single canted bond.
This excitation does not move unimpeded, 
because at each step its motion depends on the presence 
of an adjacent short bond (which only appears surreptitiously).

The MC update scheme is rather straightforward, since
each of the $\mathsf{s}_j$ terms contributing to the discrete evolution
operator, defined in Eq.~\eqref{EQ:H-shift},
maps single configurations to single configurations and
always does so with the same unit weight. The one exception is the special case 
where a canted bond with its rightmost edge at $j=N-2$
is acted upon by $\mathsf{s}_{N-1}$, which annihilates the state:
\begin{equation}
\mathsf{s}_{N-1} \ket{\po \cdots \pc \co_i \po \cdots \pc \cc_{N-2} \po \pc } = 0.
\end{equation}

As a starting point, we take a trial state $\ket{\psi_\text{trial}} = \ket{\psi_\text{mf}}$,
constructed according to the third line of Eq.~\eqref{EQ:Excited-state}.
Individual spin configurations are drawn from the distribution
defined by the wave function amplitudes. To select a spin configuration, we 
choose sites $i$ and $j$ with probability proportional to $g_{ij}$
and assign them both spin up. We then generate the three Dyck word segments that
are needed to fill in the remaining spins. 
(Only two Dyck words are needed if $j=i+1 > 2$ and only one if $i=1$ and $j=2$.)
Dyck words of a desired length are constructed using the 
biased random walk
procedure proposed in Ref.~\onlinecite{Adhikari-PRB-19}.

Formally, the trial state itself can be expressed in terms of the true eigenstates:
\begin{equation}\label{EQ:trial-wf}
\ket{\psi_\text{trial}} = a\ket{\psi_1} + b\ket{\psi_1^{\prime}} + c\ket{\psi_1^{\prime \prime}}  + \cdots
\end{equation}
Here, $\ket{\psi_1}$, an eigenstate with energy $E_1$, the lowest-lying state in the $S^z_\text{tot} = 1$ spin sector, and $\ket{\psi_1^{\prime}} , \,\ket{\psi_1^{\prime \prime}}, \ldots$ are higher energy states in the same sector with energies $E_1^{\prime} < E_1^{\prime \prime} < \cdots$.
Applying the power method gives
\begin{equation}\label{EQ:power-QMC}
\begin{split}
\mathsf{U}^n&\ket{\psi_\text{trial}} \\
&= a\mathsf{U}^n\ket{\psi_1} + b\mathsf{U}^n\ket{\psi_1^{\prime}} +  \cdots \\
&= (\mathcal{C}-2E_1)^n\Bigl[ a\ket{\psi_1} + b\,{\underbrace{\!\left(\frac{\mathcal{C}-2E_1^{\prime}}{\mathcal{C}-2E_1}\right)\!}_{< 1}}^n\,\ket{\psi_1^{\prime}}  +  \cdots  \Bigr]\\
&\approx a\mathcal{C}^n(1-2E_1/\mathcal{C})^n \ket{\psi_1} \quad \textrm{as} \; \, n \to \infty,
\end{split}
\end{equation}
in which the terms that are not proportional to $\ket{\psi_1}$
are suppressed 
because of the diagonal offset $\mathcal{C} = N-2$ and the ordering
of the energy values.
In other words, repeated application of the discrete evolution operator
causes the system to relax into the lowest energy state in the
($c=1,m=0$) sector to which $\ket{\psi_\text{trial}}$ belongs.
The positive constant $a = \overlap{\psi_\text{trial}}{\psi_1} \neq 0$, 
represents the overlap between the trial state and the true first excited state.
It is helpful numerically to have $a$ bounded well away from 0, which we
have accomplished by our choice of a good trial state.

Following the notation in Eq.~\eqref{EQ:H-shift}, the 
$n$-step evolution described in 
Eq.~\eqref{EQ:power-QMC} can be expanded to give
\begin{equation}\label{EQ:MC-sampling}
\begin{split}
\mathsf{U}^n\ket{\psi_\text{trial}}
&= \Biggl(\sum_{j=2}^{N-1}\mathsf{s}_j\Biggr)^n\ket{\psi_\text{trial}}\\
& = \sum_{j_n=2}^{N-1}\mathsf{s}_{j_n} \cdots \sum_{j_2=2}^{N-1}\mathsf{s}_{j_2}\sum_{j_1=2}^{N-1}\mathsf{s}_{j_1}\ket{\psi_\text{trial}}\\
&=\sum_{\{ j_1,\ldots,j_n\} }\!\!\mathsf{s}_{j_n} \cdots \mathsf{s}_{j_2}\mathsf{s}_{j_1}\ket{\psi_\text{trial}}.
\end{split}
\end{equation}
In our simulations, the final line in Eq.~\eqref{EQ:MC-sampling}
is evaluated by sampling over all possible operator strings
(each corresponding to a particular history of rearrangements of
the trial state). Such a calculation is asymptotically exact in the number of samples,
and the convergence is quick, since there is no sign problem.
As per Fig.~\ref{FIG:vertices}, the algorithm can be connected to the familiar language 
of world-line Monte Carlo~\cite{Assaad-Springer-08,Todo-Springer-13}.

\begin{figure}
\begin{center}
\includegraphics{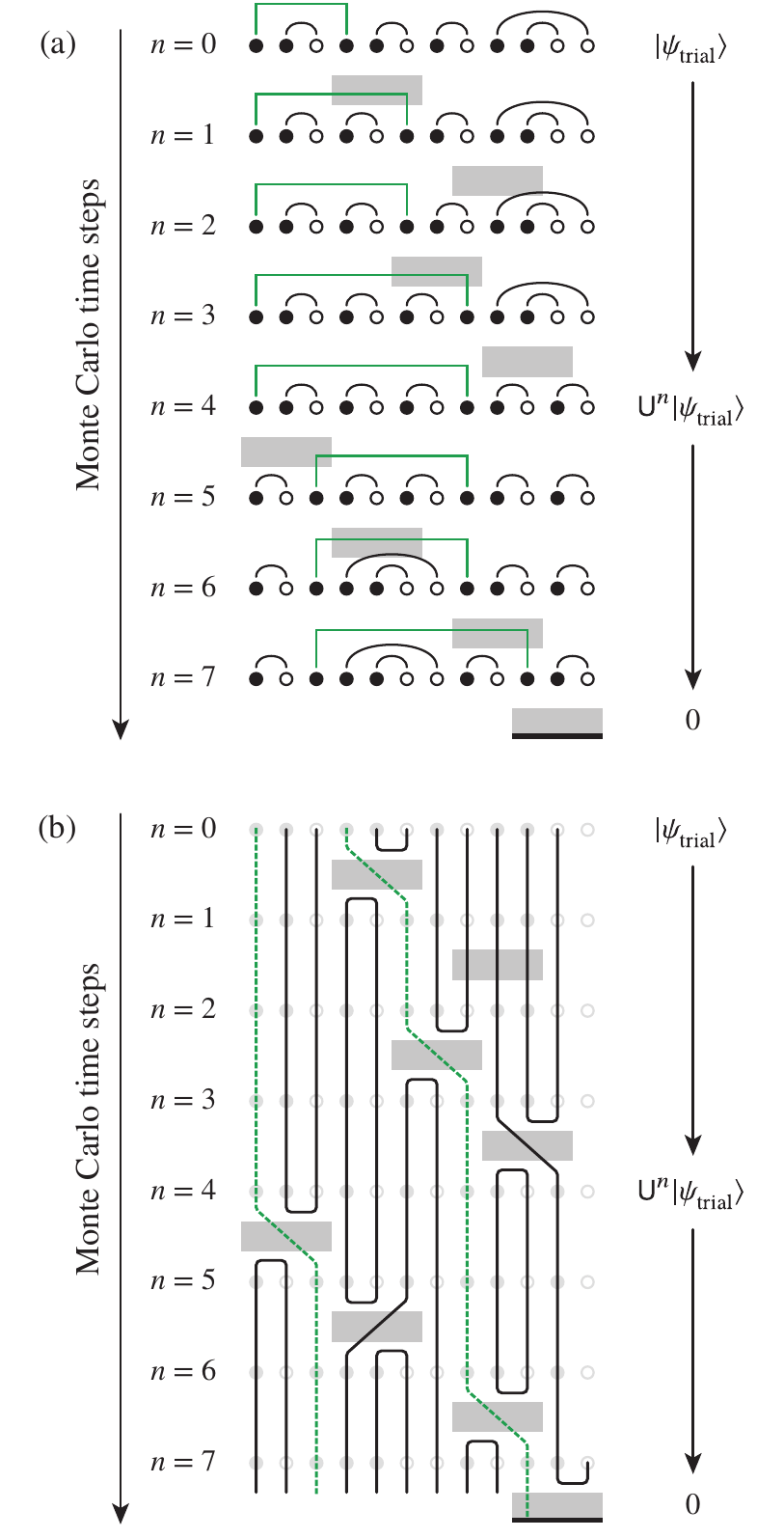}
\end{center}	 
 	 \caption[]{\label{FIG:Monte-Carlo-Update}
	 The diagram shows a snapshot of the Monte Carlo sampling for an $N=12$ system, with a progression of spin reconfigurations running from top to bottom. The lifetime of the excited state is related to the time it takes on average to traverse the system from left to right. In this particular instantiation, an excited bond is injected at $n=0$
	 and allowed to evolve until its annihilation at $n=7$ when the excited bond reaches the right edge of the chain. 
	 Each gray rectangular bar marks the application of a particular 
	 three-site $\mathsf{s}_j$ term from Eq.~\eqref{EQ:H-shift}, 
	 which amounts to a short-bond-shuffle operation,
	 as illustrated in Fig.~\ref{FIG:vertices}.
	 If the selected block does not include a short bond, then the state remains unchanged (as, e.g., in $n=1 \to 2$). Therefore, the excited bond moves only if the randomly selected block includes both a short conventional bond and one end of the excited bond.
	 An excited bond may be jammed until the background is rearranged to put a short bond adjacent to it.
	 An update of the rightmost three-site block returns a zero if it contains the right end of an excited bond.
	 The update kills the state when the right end of the bond is at $j = N-2$ and 
	 the update attempts to move 
	 it rightward.
	 }
\end{figure}

The qualitative picture is that an excited canted bond is injected at Monte Carlo time $n=0$
(with greatest likelihood of being short and of appearing toward the left edge of the chain).
The excited bond is allowed to propagate, with the bond end points $i$ and $j$ executing a kind of single-file diffusion. The bond is annihilated when $j$ finds its way to the right edge of the chain.

We argue that the survival histogram for a canted bond after $n$ projections steps must have the asymptotic form $f_n=f_0(1-\epsilon_1)^n$, with $f_0 = a\mathcal{C}^n$ and
$$\epsilon_1 = \frac{2E_1}{\mathcal{C}} = \frac{E_1}{N/2-1}.$$
Accordingly, the average excited bond lifetime is 
\begin{equation}\label{EQ:excited-lifetime}
\bar{\tau} = \frac{\sum_{n=0}^{\infty}n(1-\epsilon_1)^n}{\sum_{n=0}^{\infty}(1-\epsilon_1)^n}
 = \frac{1}{\epsilon_1}-1
 = \frac{N/2-1}{E_1}-1
 \sim N^{z+1},
\end{equation}
since $\Delta = E_1 \sim N^{-z}$.
Here, the fact that the $N-2$ terms in the Hamiltonian are sampled one at a time
contributes an extra $O(N)$ operations in the calculation of the lifetime. 
In the spirit of Eqs.~\eqref{EQ:prob2-ij} and 
\eqref{EQ:prob1-ij}, one- and two-sided measurements are performed for the observable $\hat{O}$ using 
\begin{equation}\label{EQ:meaurement}
\begin{split}
\expectation{\hat{O}}_\text{1-sided} &= \frac{\bra{R} \hat{O}\mathsf{U}^M \ket{\psi_\text{trial}}}{\bra{R} \mathsf{U}^M \ket{\psi_\text{trial}}},\\
\expectation{\hat{O}}_\text{2-sided} &= \frac{\bra{\psi_\text{trial}} \mathsf{U}^M\hat{O}\mathsf{U}^M \ket{\psi_\text{trial}}}{\bra{\psi_\text{trial}} \mathsf{U}^{2M} \ket{\psi_\text{trial}}}.
\end{split}
\end{equation}
Once again, $\ket{R}$ is a disordered reference state. The number of projection steps is $M = N^p$, where
$p$ is an exponent that should be almost as large as the dynamical exponent; we use $p=3$
throughout this work, having convinced ourselves via experimentation and careful
benchmarking that such a value is sufficient.

The one- and two-sided measurements are similar with respect to implementation.
The main difference is that the two-sided version requires two trial states (in
the bra and ket), and so we have to bias the sampling according to an additional
set of trial wave function weights. This is easily accomplished with rejection
sampling: we accept updates if  $r < g_{i\prime j^\prime}/g_{i,j}$,
 where $r \shortleftarrow [0,1)$ is a random number drawn from a uniform distribution on the unit interval; $i',j'$ are the canted bond end points {\it after} $2M$ update steps (from the proposed operator string) have been applied, and $i,j$ are the end points {\it before} any update steps have been applied.

An example operator string is shown in Fig.~\ref{FIG:Monte-Carlo-Update}. 
The upper panel shows the sequence of bond reconfigurations,
and the lower shows the corresponding world-line representation.

\section{Results and discussion}
\label{SECT:Results}

\begin{figure}
\begin{center}
    \includegraphics[width=0.9\columnwidth]{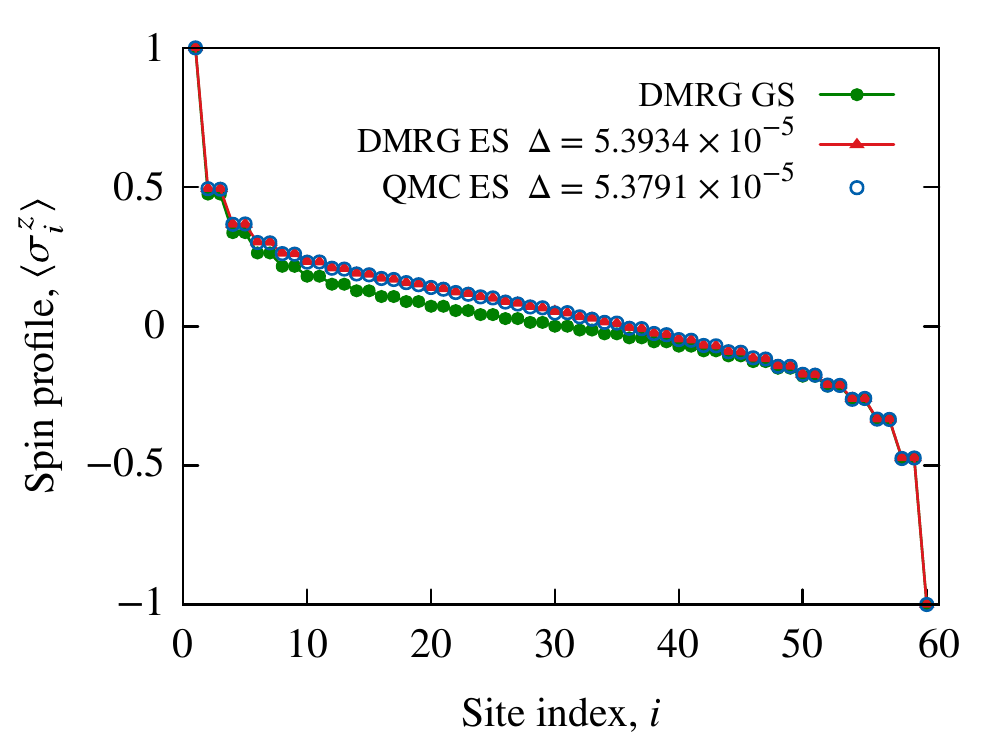}
\end{center}    
    \caption{\label{FIG:spin-profile}
    The spin profile is the expectation value of the spins projected along the z-axis, spatially resolved over lattice sites $i=1,2,\ldots, N$. Results are given here for system size $N=60$. GS (ES) denotes the expectation values computed in the ground (excited) state wave function. QMC results are measured after $M = N^3$ projections on the trial state. These
compare favorably to the results from DMRG.
The slightly lower energy of the QMC result suggests that the QMC wave function is superior
in the variational sense.
   }    
\end{figure}

As a test of our numerics, we have confirmed that the spin profile measured with our 
quantum Monte Carlo (QMC) is in agreement with results from DMRG calculations~\cite{ITensor}. 
See Fig.~\ref{FIG:spin-profile}. Computing the difference between the excited- and ground-state 
spin profiles, we see that the spin excess (corresponding to the average position of the physical spin 
flip at $j$) is localized on the left side of the chain, with a peak near $N/3$. This is 
consistent with the mean-field picture that a bond $\co_i \cdots \cc_j$ feels an attraction 
toward the left edge that works to minimize the value of the index $i$ and a mutual attraction 
between the bond end points that works to minimize the bond length $j-i$. Alternatively, we might
say that the spin excess is largest on the left because the right
edge of the chain is a sink for the excited bond probability.

\begin{figure}
\begin{center}
    \includegraphics[width=0.9\columnwidth]{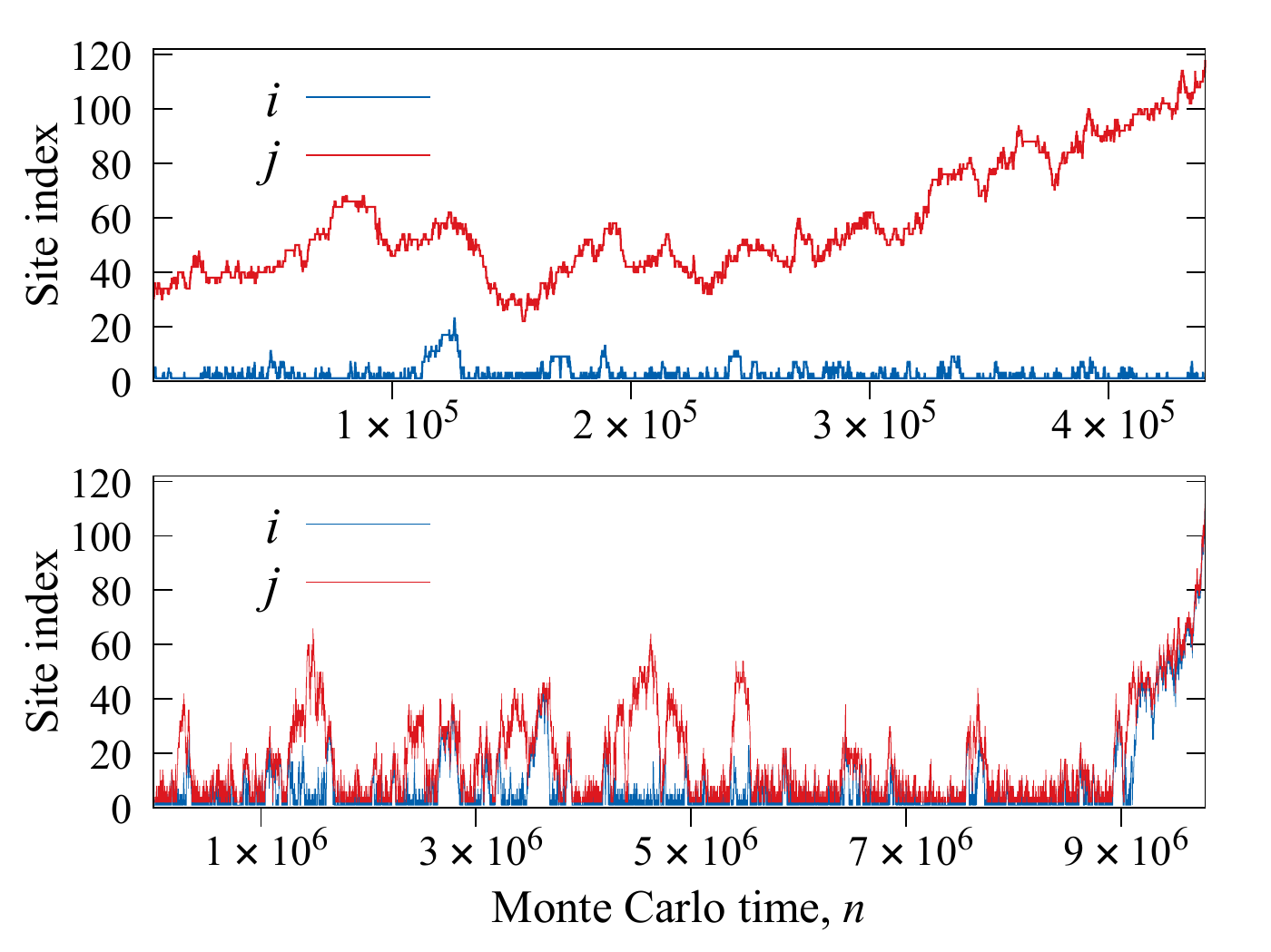}    
\end{center}
    \caption{\label{FIG:trajectory-sample}
    Each of the two panels shows the randomly generated history of a 
    canted bond $\co_i \cdots \cc_j$ evolving
    in MC time for the $N=120$ Fredkin spin chain.
    The traces show the paths of the left ($i$, blue) and right ($j$, red)
    bond end points.
    The upper panel highlights the somewhat rare behavior
    in which the bond reaches across the entire system;
    in that case, the time to traverse the sample is much faster
    than average ($\tau \ll \bar{\tau} \approx 1 \times 10^7$).
    The lower panel shows the more typical behavior
    in which the bond migrates across the system while remaining short.
    }
\end{figure}

\begin{figure}
\centering
    \includegraphics[width=0.9\columnwidth]{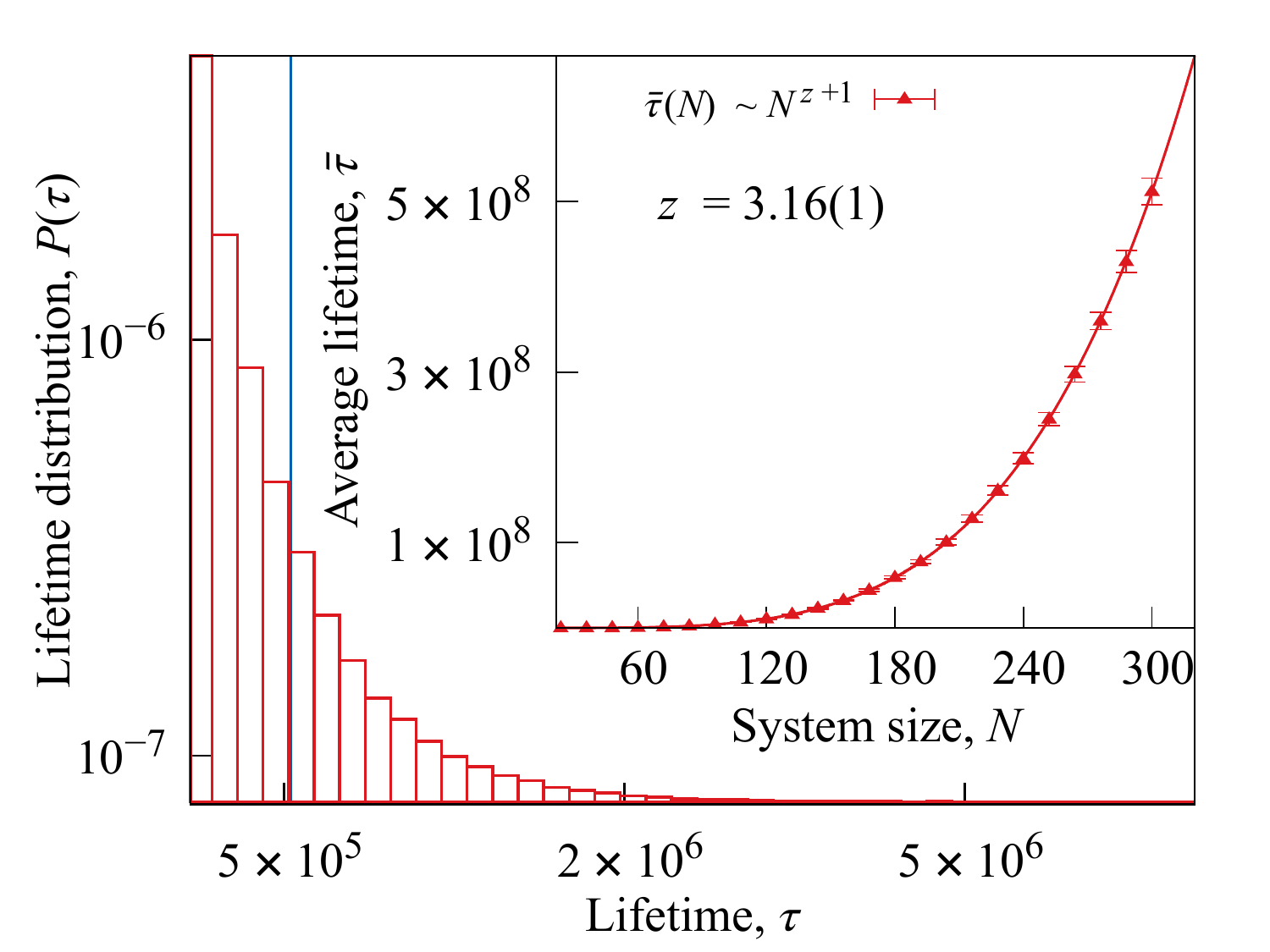}
    \caption{\label{FIG:lifetime}
    Main panel: The lifetime distribution $P(\tau)$ of a canted bond in the Fredkin spin chain
    is presented for a single system size.
    The vertical blue line marks the mean lifetime, $\bar{\tau} = \int\,d\tau P(\tau)$.
    The lifetime histogram is non-normal (heavily right-skewed) and dominated by a long tail with significant
    weight out to many times $\bar{\tau}$.
    Inset: The mean lifetime is plotted for various system sizes $N = 24,36, \ldots, 288, 300$.
    Each measured lifetime is the first-passage time for the excited bond to reach the right edge of the chain.
    The average is taken over many bond-trajectory histories.
    The best-fit line passing through the data is a powerlaw
    with exponent $1+z = 1 + 3.16(1)$.}
   \end{figure}

Figure~\ref{FIG:gplot} shows that the most probable excited bond has indices $i=1,j=2$. In this configuration $\mathcal{D}'$ and $\mathcal{D}''$ are null; $\mathcal{D}'''$ achieves its maximum length, $N-2$; and the probability of finding a short planar bond at site 3 and 4 is at its lowest.
As a result, the excited bond spends most of the time quasi-bound at the left end of the chain, making only rare excursions to the right (see Fig.~\ref{FIG:trajectory-sample}). 
The bond lifetime
depends on the statistics of escape events in which the excited bond breaks free of the left edge and completes a transit across
the spin chain. These transits fall into two main classes, long-lived trajectories in which the moving bond remains short [with $j-i = O(1)$] and short-lived trajectories in which $i$ remains pinned to the left edge of the chain while the bond stretches to system-spanning size [$j-i = O(N)$].
Hence the distribution of lifetimes is highly non-Gaussian, with a long tail
extending well below the main peak. (See the main panel of Fig.~\ref{FIG:lifetime}.)
A four-parameter fit of the average lifetime to the form $\bar{\tau} = (u_0 + u_1/N)N^{1+z+\varv_1/N}$
gives an estimate of the dynamical exponent $z=3.16(1)$.

The rationale for this fitting form is that the deeply asymptotic behavior $\bar{\tau} = u N^{1+z}$ depends on two parameters: an overall multiplicative constant and the dynamical exponent $z$. For system sizes that are only moderately large, we expect both $u\to u(N)$ and $z\to z(N)$ to take on effect size-dependent values. The form we have chosen, $u(N) = u_0 + u_1/N$
and $z(N) = z_0 + z_1/N$, assumes that each parameter keeps only the first subleading corrections. Admittedly, other choices are possible. However, additional corrections at $O(N^{-2})$ or beyond offer too many free parameters and run the risk of overfitting, whereas throwing out the subleading corrections entirely is unworkable, because we are too far from the deeply asymptotic limit (which sets in at thousands or tens of thousands of sites for this model).
An alternative is to fit to $u N^{1+z}$ in narrow bands of similarly sized simulations and then to extrapolate the size-dependent $u$ and $z$ values to $N\to\infty$. This is the style of analysis used by Chen and coworkers in Ref.~\onlinecite{Chen-JPA-17}. We have confirmed that the values of $z$ we extract from our own simulation data are consistent across various approaches.

\begin{figure}
    \centering
    \includegraphics[width=0.9\columnwidth]{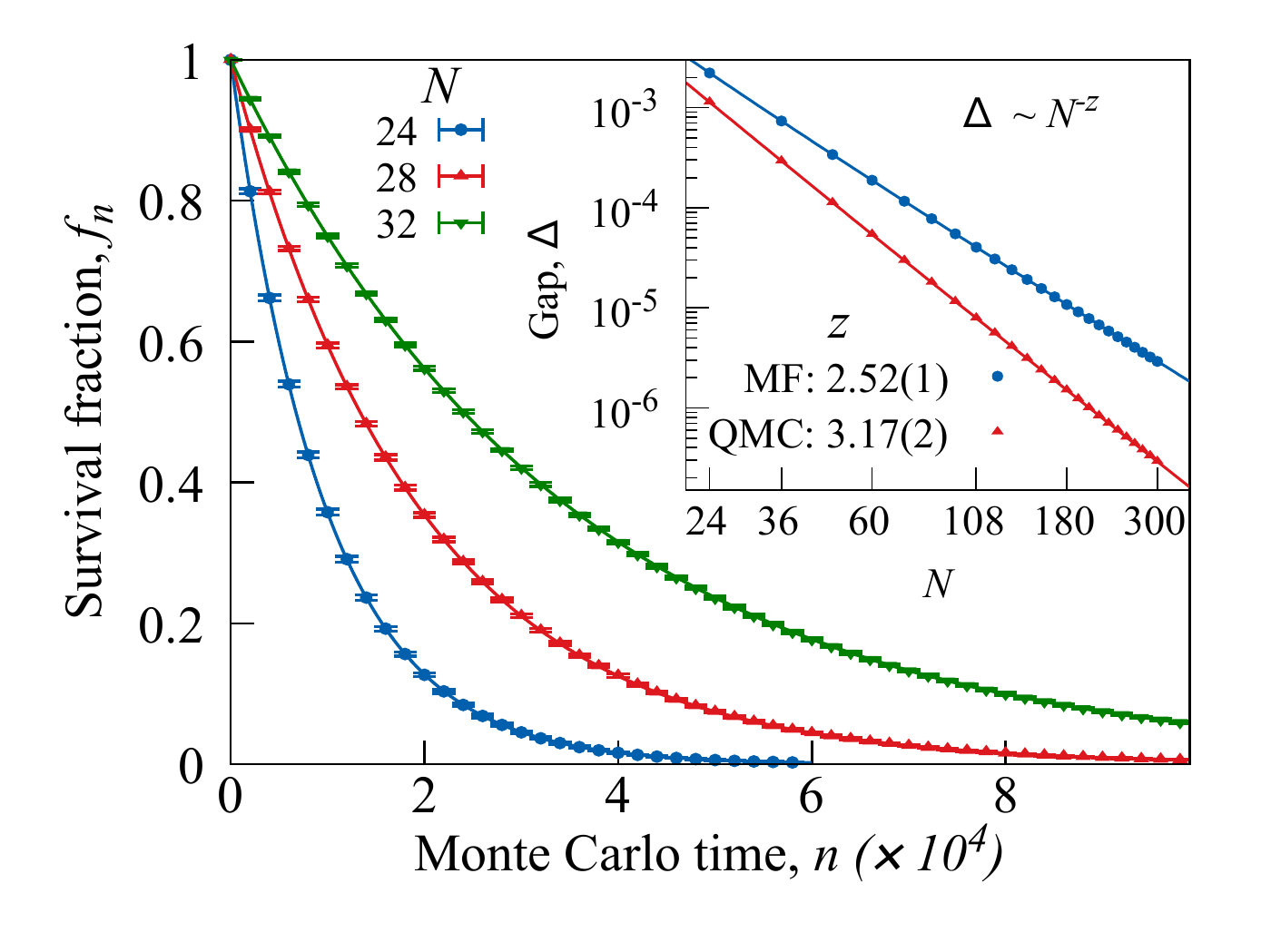}
    \caption{\label{FIG:survival-fraction}
    In the main panel, survival fraction versus Monte Carlo time (measured after an initial $N^p=N^3$ projection steps) is plotted. The three lines show the data fitted to the asymptotic form $f_n=f_0(1-\epsilon_1)^n$ with normalized $f_0 = 1$ and $\epsilon_1 = \Delta/(N/2-1)$.
        The inset shows fits of the form
        $\Delta(N) = \bigl(u_0 + u_1/N\bigr)N^{-z-\varv_1/N}$ running through each set of gap data extracted from the survival fraction for finite system sizes $N = 24,36, \ldots, 288, 300$.
}
\end{figure}

As a further check, we consider an alternative analysis.
The survival histogram in Fig.~\ref{FIG:survival-fraction} is a plot of the fraction of states that survive up to a given Monte Carlo time. Here, observation begins after $M = N^3$ initial projections on the trial state, which marks time zero and the renormalization of the survival fraction to 1. As the system size increases, the states survive longer. Since the Fredkin has a zero-energy ground state, the excited state energy can be extracted directly from the fitting form $f_n=f_0(1-\epsilon_1)^n$. 
We carry out such a fit for a dense grid of system sizes up to $N=300$.
Applying a finite-size scaling ansatz $E_1(N) = \Delta(N) = (u_0 + u_1/N)N^{-z-\varv_1/N}$ 
then produces $z=3.17(2)$.
This estimate has a slightly larger uncertainty but is consistent with the previous value
of $z=3.16(1)$ extracted directly from the quantum dynamics.

\section{Conclusions}
\label{SECT:Conclusions}

We formulated a sign-problem-free Monte Carlo scheme for
the Fredkin spin chain, organized within the 
framework of a bond representation of the model's Hilbert space.
We have established a convention such that
each possible spin configuration admits a unique rewriting 
as a product of planar, mismatch, and canted bonds.
Our numerical implementation is targeted to the specific case 
of no mismatch bonds and one canted bond ($m\!=\!0,c\!=\!1$),
but it is straightforward to generalize the algorithm to arbitrary numbers of either bond type.

Unlike level-spectroscopic approaches that infer the dynamics
from the finite-size scaling of states in the energy spectrum, our approach
relies on direct observation of the quantum dynamics.
We simulate the low-lying $\sztot = 1$ excited state
by way of the following numerical experiment.
An up-canted bond is injected toward the left edge of the spin chain.
It executes a  biased diffusive walk across the finite chain segment,
moving through the background of conventional bonds
subject to the activation rules for hopping that require
short-bond adjacency. Finally the canted bond is annihilated 
when its own right end point (the location of the physical
spin flip) reaches the right edge of the chain.
Our simulations are stochastic in nature.
We generate many long sequences of bond of rearrangements 
and average over those histories to obtain
statistics about the excited bond's lifetime.
Our best estimate of the dynamical exponent is $z \doteq 3.16(1)$.

We remark that the technique we have developed can equally 
be used to studying the low-lying $\sztot=0$ excitation---or in any sector 
with quantum numbers $(c,m$). In the $\sztot=1$ case that we have presented, a 
single {\it canted} bond $(c=1,m=0)$ is injected according to a distribution 
skewed toward the left edge of the system and allowed to execute a 
quantum random walk until it annihilates at the right edge. In the $\sztot=0$ 
case, a single {\it mismatch} bond $(c=0,m=1$) is injected toward the 
{\it middle} of the spin chain and made to evolve until it annihilates 
at {\it either the left or right edge}. In that case,
the gap closes somewhat slower, with a dynamical exponent $z \approx 2.7$.
In either case, the implementation of the QMC simulation is nearly identical. 
In this paper, we made the choice to focus on the $\sztot=1$ sector, since it 
corresponds to the lowest-lying excitation and is the one that dominates the 
long-time dynamics.

We have also presented a mean-field analysis in which 
the Dyck word segments between the excited bond
end points are imagined to be always in a Dyck-word 
ground-state configuration. In this picture, the excited bond
executes a biased diffusive walk across
the spin chain, but its motion is slow
because it must drag itself through the molasses
of the spin background.
The underlying
assumption is that the time-scale for relaxing
back to a Dyck word is much faster than that of 
the excited bond's motion---so that the background
spins view the excited bond as frozen in place
and the excited bond experiences the spin background spins as 
a memory-less effective medium.

The mean-field calculation yields a value
of $z_\text{mf} \approx 5/2$.
We understand this to be a lower bound on the true 
dynamical exponent, since the processes
neglected in the mean-field ansatz are
ones that only further slow the motion.
The mean-field result, despite its simplicity, 
is already nontrivial 
because it predicts motion slower 
than the $z=2$ dynamics typical of ferromagnets.

The structure of the mean-field wave function 
helps to explain the behavior observed in the full QMC simulations.
It suggests that an up-canted bond with end points
at $i$ and $j$ feels a dual attraction,
with both $i$ and $j-i$ tending to small values.
The excitation prefers to be pinned
to the left edge and to be as short as possible.
Our simulations confirm that there are two main pathways
for $j$ to wander across the system en route to
annihilation. Either the short bond
breaks free from the left edge
of the chain and remains short while traveling
or the left end point remains pinned
at $i=1$ and $j$ moves toward the right edge
of the chain by stretching the excited bond 
across the whole system.

\end{document}